\documentclass[conference]{IEEEtran}
\usepackage{mathrsfs}
\usepackage{amsfonts}
\usepackage{ifpdf}
\usepackage{cite}
\ifCLASSINFOpdf
\usepackage[pdftex]{graphicx}
\else \fi
\usepackage[cmex10]{amsmath}
\usepackage{array}
\usepackage{mdwmath}
\usepackage{mdwtab}
\usepackage{eqparbox}
\usepackage[tight,footnotesize]{subfigure}
\usepackage{algorithmic}
\usepackage{algorithm}
\usepackage{amsmath}
\usepackage{epsfig}
\usepackage{ae}
\usepackage{amssymb}
\usepackage{times}
\usepackage{amsmath}   % add1
\usepackage{booktabs}  % add2
\usepackage{threeparttable} %add3
\usepackage{multirow}       %add4
\usepackage{xcolor}

\newtheorem{theorem}{\textbf{Theorem}}
\newtheorem{lemma}{\textbf{Lemma}}

\begin{document}
%----------------------------------------Make Title -----------------------------------------

\title{On Peak Age of Information in Data Preprocessing enabled IoT Networks}

%\title{Service Selection and Resource Allocation in Multi-tier Heterogeneous Cellular Networks: A load balancing perspective}

%-------------------------------------------------------------------------------------------
\author{\IEEEauthorblockN{Chao Xu$^{\dag}$, Howard H. Yang$^{\ddag}$, Xijun Wang$^*$, and Tony Q. S. Quek$^{\ddag}$}
\IEEEauthorblockA{$^{\dag}$School of Information Engineering, Northwest A\&F University, Yangling, Shaanxi, China\\
$^{\ddag}$Information System Technology and Design Pillar, Singapore University of Technology and Design, Singapore \\
$^*$School of Electronics and Communication Engineering, Sun Yat-sen University, Guangzhou, China}
\thanks{
This paper is supported by National Natural Science Foundation of China (61701372) and Talents Special Foundation of Northwest A\&F University (Z111021801).
}}

%\author{
%Chao Xu, {\em Member, IEEE}, Howard H. Yang, {\em Member, IEEE}, Xijun Wang, {\em Member, IEEE}, \\ and Tony Q. S. Quek, {\em Fellow, IEEE}
%%\author{Chao Xu, {\em Member, IEEE}, Min Sheng, {\em Member, IEEE}, Vineeth S. Varma, \\ Tony Q. S. Quek, {\em Senior Member, IEEE}, and Jiandong Li, {\em Senior Member, IEEE}

%%
%%C. Xu is with the School of Information Engineering, Northwest A\&F University, Yangling, Shaanxi, China (e-mail: cxu@nwafu.edu.cn).
%%
%%X. Wang is with State Key Laboratory of ISN, Xidian University, Xi`an, China (e-mail: xijunwang@xidian.edu.cn).
%%
%%H. H. Yang and T. Q. S. Quek are with the Singapore University of Technology and Design, Singapore 487372 (e-mail: hao\_yang@mymail.sutd.edu.sg; tonyquek@sutd.edu.sg).
%%
%%
%%}
%}

\maketitle
%---------------------------------------Make Abstract--------------------------------------
\begin{abstract}
Internet of Things (IoT) has been emerging as one of the use cases permeating our daily lives in 5th Generation wireless networks, where status update packages are usually required to be timely delivered for many IoT based intelligent applications. Enabling the collected raw data to be preprocessed before transmitted to the destination can provider users with
better context-aware services and lighten the transmission burden. However, the effect from data preprocessing on the overall information freshness is an essential yet unrevealed issue. In this work we study the joint effects of data preprocessing and transmission procedures on information freshness measured by peak age of information (PAoI). Particularity, we formulate the considered multi-source preprocessing and transmission enabled IoT system
as a tandem queue where a priority M/G/1 queue is followed by a G/G/1 queue. Then, we respectively derive the closed-form and an information theoretic approximation of the expectation of waiting time for the formulated processing queue and transmission queue, and further get the analytical expressions of the average PAoI for packages from different sources. Finally, the accuracy of our analysis is verified with simulation results.

\end{abstract}

%--------------------------------------------- Make Key words-----------------------------

\section{Introduction}

As one of the promising technologies for the 5th Generation (5G) wireless networks, Internet of Things (IoT) has attracted significant attention from both academia and industry in recent years \cite{IoT_5G_2017,IoT_3gpp}. With the help of IoT, devices can sense and even interact with the physical surrounding environment, thereby providing us with many valuable and remarkable context-aware applications to improve the quality of our lives at an efficient cost \cite{Survey_IoT_Applications_2015}.
Limited by the resource for data transmission, the collected raw data in IoT networks are usually preprocessed before transmitted to the final destination (e.g., actuator or monitor) to provide users with better context-aware services\cite{IoT_Context_Aware_2018}, which may be enabled, for instance, by resorting to the new emerging edge/fog computing technology \cite{MEC_IoT_2018}. Wherein, many IoT based intelligent applications, including the automatic control of electric appliance \cite{Conference_Transit_Data}, intelligent transportation network \cite{Traffic_IoT}, and event monitoring and predication for health safety \cite{Data_Mining_IoT}, require fresh information for devices to response with adequate action.

In order to quantify the level of ``freshness'' from the information delivered, age of information (AoI) \cite{AoI_Org_2012} and peak age of information (PAoI) \cite{PAoI_2014_ISIT}  have been recently introduced. Particularly, AoI measures the time elapsed since the latest received update package was generated, while PAoI provides information about the maximum value of AoI for each update. Since PAoI captures the extent to which the update information is stale, it has been regarded as an efficient new metric to investigate the freshness of the delivered information in IoT networks with a single data source \cite{PAoI_2014_ISIT,PAoI_TIT_2016} and multiple data sources \cite{Multi_Source_ISIT,Multi_SOurce_Preemptive_2018,Max_Min_PAoI}. Considering the system with a single source and one destination, authors in \cite{PAoI_2014_ISIT} and its journal version \cite{PAoI_TIT_2016} analyzed the effects of different data management policies (discard or replacement in the buffer) on PAoI by modeling distinct queueing models, and meanwhile analyzed their effectiveness on performance improvement in different scenarios. Focusing the IoT networks with multiple data sources, work \cite{Multi_Source_ISIT} analyzed the system performance by considering general service time distributions, and tried to optimizing the packages arrival rates to minimize its defined average PAoI-related system cost. In \cite{Multi_SOurce_Preemptive_2018} the serving policies of preemption and package discard were allowed for the first-come-first-served (FCFS) based transmission when the transmitter was busy, and expressions of the average PAoI were derived. Authors in \cite{Max_Min_PAoI} considered the interactions among distinct transmission links and proposed link scheduling algorithms to minimize the maximum PAoI of packages from different sources.

While good studies on PAoI have been presented for IoT networks with multiple data sources \cite{Multi_Source_ISIT,Multi_SOurce_Preemptive_2018,Max_Min_PAoI}, these work treated the data aggregator purely as a transmitter and thus do not apply to cases where the data shall be preprocessed (e.g., data compression and aggregation) to filter out the redundancy or even extract the ``intrinsic content'' from collected raw data before any transmission procedure begins. As such, it calls for a focus on investigating the joint effect of data preprocessing and transmission on PAoI.
The most related work to this topic comes from \cite{PAoI_Camera_Networks}, which studied the wireless camera networks consisting of multiple sources and fog nodes, and proposed a modular optimization algorithm to minimize the achieved maximum PAoI by optimally assigning processing nodes and scheduling transmission links. However, the effect of the processing procedure (e.g., processing policy and time) and update package arrival rate on information freshness has not been jointly investigated in \cite{PAoI_Camera_Networks} nor, to the best of our knowledge, in other existing researches.

In this paper, we consider an IoT network that consists of a data aggregator and a destination. The aggregator will first preprocess the status update packages generated from multiple sources with different priorities and then forward the processed data to the destination node (e.g., actuator or monitor) via a wireless channel according to the FCFS discipline. We establish a tandem queue to model the joint effect from data preprocessing and transmission, which is general and captures all the key features in an IoT network, including the prioritized data processing, queueing, and wireless channel fading. Moreover, we respectively derive a closed-form expression and an information theoretic approximation of the expectation of waiting time for the processing queue and transmission queue, and further get the analytical expressions of the average PAoI for packages from different sources. Simulation results show that it's plausible to qualitatively capture the variation trend of the average PAoI with our theoretical analysis, when the processing time is dominating or comparable with the transmission time, e.g., for computation intensive applications.

\section{System Model}
\begin{figure} [!t]
\centering
\leavevmode \epsfxsize=3.0in  \epsfbox{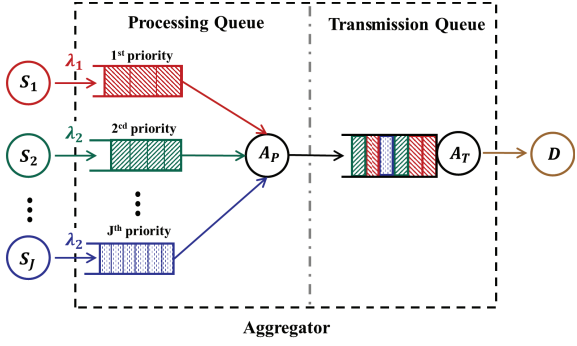}
\centering \caption{Illustration of the tandem queueing model for the considered IoT network.} \label{Fig:Fig1}
\end{figure}
\subsection{Network Model}
We consider an IoT system which consists of $J$ data sources, denoted by $\mathcal{S} = \{S_1, S_2, \cdots, S_J\}$, a data aggregator that is able to perform data preprocessing as well as transmission, and a destination node, as depicted in Fig.~\ref{Fig:Fig1}. Each source keeps collecting information from the ambient environment and periodically updates the status to the aggregator, whereas the data packages from source $S_j$ arrive at the aggregator according to an independent Poisson distribution with parameter $\lambda_j, \forall j \in \mathcal J = \{1, 2, \cdots, J\}$.
Upon receiving the status information, the aggregator conducts a preprocessing to the data packages before forwarding them to the destination node.
As significance of information from each source can vary with respect to the content, we allow the aggregator to process the incoming packages with different priorities.
Without loss of generality, we assume the data from $S_i$ has a higher priority than that from $S_j$ if $i \leq j$. In this regard, a generic data package can only be processed if there is no data package with a higher priority waiting to be processed. For an income data package with the $j$-th priority, we denote $C_j$ and $\tilde{C}_j$ ($\tilde{C}_j<C_j$) as the size before and after data processing, respectively, and $\frac{C_j-\tilde{C}_j}{\tau_j}$ represents the corresponding processing time, where $\tau_j$ is the equivalent processing rate of the aggregator related to the specific operation made on the data. As such, the preprocessing subsystem is formulated as a priority M/G/1 queue where the size of buffer is infinite.\footnote{We note that the following analysis also holds when we consider another function mapping from each $(C_j, \tilde{C}_j)$ to a positive real number (i.e., the processing time), since a priority M/G/1 queue can also be formulated in that scenario.}

Once a data package finishes preprocessing, it will be pushed into an infinite-size queue at the transmitter according to the first-come-first-served (FCFS) discipline. We term this buffer the transmission queue. At the transmitter side, we consider each package is sent with constant power $p_A$, and the propagation channel is subjected to small scale Rayleigh fading with unit mean and large scale path loss that follows power law, with path loss exponent $\alpha > 2$. { In addition, the spectrum for data transmission occupies a bandwidth of $B$ Hz.}

\begin{figure}[!t]
\centering \leavevmode \epsfxsize=3.0in  \epsfbox{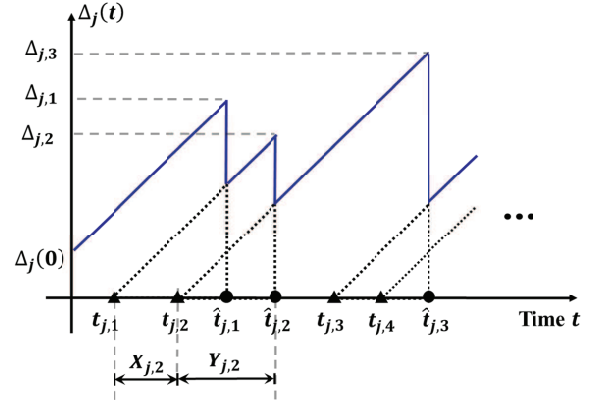}
\centering \caption{Example of the AoI evolution process for the $j$-th source at the destination node. The time instant of package arrival at the aggregator and the destination node are marked as $\blacktriangle$ and $\bullet$, respectively.} \label{Fig:Age_variation_Example}
\end{figure}
\subsection{Age of Information}
We denote $t_{j,n}$ the time instant when the $n$-th package from $S_j$ arriving at the aggregator, and denote with $\hat{t}_{j,n}$ the time instant that this package arrived at the destination node. Meanwhile, the age of information (AoI) of source $S_j$ is defined as ${\Delta _j}( t ) = t - {u_j}( t )$, where ${u_j}( t )$ is the generation time of the most recently received package from $S_j$ until time instant $t$ \cite{AoI_Org_2012,PAoI_2014_ISIT,PAoI_TIT_2016,Multi_Source_ISIT,Multi_SOurce_Preemptive_2018,Max_Min_PAoI}.
{An example of the AoI evolution process ${\Delta _j}(t)$ for the $j$-th source is illustrated in Fig. \ref{Fig:Age_variation_Example}.\footnote{Similar as previous studies \cite{Multi_Source_ISIT,Multi_SOurce_Preemptive_2018}, we
consider the time spent on the transmission from sources to the aggregator negligible since they are generally integrated as a complete system and connected via high speed wired links.}}
It can be seen that the $n$-th peak value of ${\Delta _j}(t)$ is achieved just before the $n$-th package arrives at the destination node, which is defined as the peak age of information (PAoI) corresponding to the $(n-1)$-th received package and denoted by $\Delta_{j,n}$, i.e.,
\begin{align} \label{Eq:Peak_Age_Information_Instance}
\Delta_{j,n}=
\begin{cases}
{\Delta _j}\left( 0 \right)+\hat{t}_{j,n}, &n=1 \\
X_{j,n}+Y_{j,n}, & n >1
\end{cases}
\end{align}
where ${\Delta _j}( 0 )$ denotes the initial age of the last received data at the start time, $X_{j,n}$ represents the time interval between $t_{j,n}$ and $t_{j,(n-1)}$, and $Y_{j,n}$ represents the time interval between $\hat{t}_{j,n}$ and $t_{j,n}$, i.e., $X_{j,n}= t_{j,n}-t_{j,(n-1)}$ and $Y_{j,n}=\hat{t}_{j,n}-t_{j,n}$. Next, we  provide detailed analyses for the joint effects of the preprocessing and transmission procedures on the achieved average PAoI for packages from different sources.

\section{Average Peak Age of Information} \label{Sec:PAoI_Analysis}
To start with, the following lemma presents a general form of the average PAoI for each source.
\begin{lemma}
The average PAoI attained for source $S_j$ is
\begin{align}\label{Eq:General_PAoI}
{\Delta _j} = \frac{1}{{{\lambda _j}}} + \tau_j\frac{C_j-\tilde{C}_j}{r} + \mathbb E \big[ {W_j^P} \big]+ \mathbb E \big[ {Z_j^T} \big] + \mathbb E \big[ {W_j^T} \big]
\end{align}
where $\mathbb E[W_j^P]$, $\mathbb E[{Z_j^T}]$, and $\mathbb E[{W_j^T}] $ represent the expected time spent in the preprocessing queue, the transmission stage, and the transmission queue, respectively.
\end{lemma}
\begin{IEEEproof}
By ergodicity, the average PAoI for source $S_j$ can be calculated as
\begin{align} \label{Eq:Average_PAOI}
{\Delta _j} &= \mathop {\lim }\limits_{t \to \infty } \frac{1}{{{N_j}\left( t \right)}} \!\! \left( {{\Delta _j}\left( 0 \right) + \hat{t}_{j,1}} +\! \sum\nolimits_{n = 2}^{{N_j}\left( t \right)} {\left( {{X_{j,n}} + {Y_{j,n}}} \right)} \right) \nonumber \\
&\mathop  = \limits^{\left( a \right)} \mathbb E \left[ {{X_j} + {Y_j}} \right] = \mathbb E \big[ {{X_j}} \big] + \mathbb E \big[ {Y_j^P} \big] + \mathbb E \big[ {Y_j^T} \big]
\end{align}
where ${{N_j}( t )}$ denotes the number of concerned packages until time instant $t$, $Y_{j,n}$ is the sum of the time a package spent in the processing subsystem $Y_{j,n}^P$ and that in the transmission stage $Y_{j,n}^T$, and ($a$) follows from the fact that the effect of $F_{j,1}={{\Delta _j}(0) + \hat{t}_{j,1}}$ vanishes as $t$ goes to infinity.

Recalling that the package arrival from source $S_j$ follows exponential distribution with parameter $\lambda_j$, we thus have $\mathbb E\big[ {{X_j}} \big]={1}/{\lambda_j}$. Moreover, for each package, the sojourn time spent in the aggregator consists of the queueing time and serving time. Hence, Eq. (\ref{Eq:Average_PAOI}) can be written as
\begin{align} \label{Eq:Average_PAOI_Detail}
{\Delta _j} &= {1}/{{{\lambda _j}}}+ \mathbb E \big[ {Z_j^P} \big] + \mathbb E \big[ {W_j^P} \big] + \mathbb E \big[ {Z_j^T} \big] + \mathbb E \big[ {W_j^T} \big]\\ \nonumber
&= {1}/{{{\lambda _j}}} + ({C_j-\tilde{C}_j})/{\tau_j} + \mathbb E \big[ {W_j^P} \big] + \mathbb E \big[ {Z_j^T} \big] + \mathbb E \big[ {W_j^T} \big]
\end{align}
where $\mathbb E [ {Z_j^P} ]$ is the average time spent for data preprocessing and is given as $\mathbb E [ {Z_j^P} ] ={(C_j-\tilde{C}_j)}/{\tau_j }$.
\end{IEEEproof}

In the following, we detail the analysis to each individual elements, i.e., $\mathbb E [ {W_j^P} ]$, $\mathbb E [ {Z_j^T} ]$, and $\mathbb E [ {W_j^T} ]$, in \eqref{Eq:General_PAoI}.
% , \textcolor{blue}{which are the expectation of the queueing time in processing subsystem, that of the serving time in processing subsystem, that of the queueing time in transmission subsystem, and that of the serving time in transmission subsystem, respectively}.

\subsection{Computing the Expectation $\mathbb E \big[ {W_j^P} \big] $}
Due to prioritized processing, a newly arriving package with priority $j$ will have to wait till the completion of data processing for the following packages:
\begin{enumerate} []
\item The package that is currently occupying the processor.
\item The packages with priorities from 1 to $j$ in the processing queue when the package arrives.
\item The packages with priorities from 1 to $j-1$ that arrive while the typical package is waiting for its service.
\end{enumerate}

We denote by $P_{A_P,B}$ the probability that the processor is busy. Using the Little's law \cite{Queueing_Performance}, we have the following\footnote{Note that Little's Law makes no assumptions on the concerned system except for the ergodicity which is a common assumption for stable systems.}
\begin{align} \label{Eq:Busy_Probability_AP}
P_{A_P,B}  = \sum\nolimits_{j = 1}^{J}  {{\lambda _j} \mathbb E \big[ {Z_j^P} \big] }  = \sum\nolimits_{j = 1}^{J}  {{\lambda _j}\frac{{{C_j} - {{\tilde C}_j}}}{{\tau _j}}}.
\end{align}
Then, we can derive $\mathbb E [ {W_j^P} ]$ as shown in Theorem \ref{Theorem:Exp_Waiting_Time_PQ}.

\begin{theorem} \label{Theorem:Exp_Waiting_Time_PQ}
For packages from source $S_j$, the expected waiting time in the processing queue is given by
\begin{align} \label{Eq:Remaining Processing_Time_Fin}
\mathbb E \big[ {W_j^P} \big] =
\frac{{\sum_{j = 1}^J {{{{{ {{\rho _j}} }^2}}}/{{{\lambda _j}}}} }}{{2\left( {1 - \sum_{i = 1}^j {{\rho _i}} } \right) \! \left( {1 - \chi_{ \{ j > 1 \} } \sum_{i = 1}^{j - 1} {{\rho _i}} } \right)}}
\end{align}
where $\rho _j={{\lambda _j} \mathbb E [{Z_j^P}] } $ denotes the load contributed by the data packages from source $S_j$, and $\chi_{ \{ \cdot\} }$ is the indicator function.
\end{theorem}
\begin{IEEEproof}
The proof is given in Appendix \ref{Pro:Exp_Waiting_Time_PQ}.
\end{IEEEproof}

\begin{figure*}
\begin{align}   \label{Eq:Waiiting_Time_Transmission_Q_Ratio}
\mu _j
% = \frac{\mathbb E\left( {W_j^T} \right)}{\mathbb E\left( {W_1^T} \right)} &\approx \frac{\lambda _j\left({{P_{{A_P},B}}\mathbb E\left( {{Z^T}} \right){\rm{ + }}\sum\limits_{i = 1}^j {{\lambda _i}\mathbb E\left( {W_i^P} \right)\mathbb E\left( {Z_i^T} \right)}  + \sum\limits_{i = 1}^{j - 1} {{\lambda _i}\mathbb E\left( {W_j^P} \right)E\left( {Z_i^T} \right)} }\right)}{\lambda _1\left({{P_{{A_P},B}}\mathbb E\left( {{Z^T}} \right) + {\lambda _1}\mathbb E\left( {W_1^P} \right)\mathbb E\left( {Z_1^T} \right)}\right)} \\ \nonumber
\!\approx \!{\lambda _j\!\left(\!{{P_{{A_P},B}}\mathbb E \big[ {{Z^T}} \big]  \!\!+\! {\lambda _j}\mathbb E \big[ {W_j^P} \big] \mathbb E \big[ {Z_j^T} \big]  \!  \!+  \! \!\sum\nolimits_{i = 1}^{j - 1} {{\lambda _i}\left(\! {\mathbb E \big[ {W_i^P} \big] \! +  \!\mathbb E \big[ {W_j^P} \big] } \!\right)\mathbb E \big[ {Z_i^T} \big] } }\!\right)}/{\lambda _1({{P_{{A_P},B}}\mathbb E \big[ {{Z^T}} \big]  \!+  \!{\lambda _1}\mathbb E \big[ {W_1^P} \big] \mathbb E \big[ {Z_1^T} \big] })}
\end{align}
\hrulefill
\end{figure*}

\subsection{Computing the Expectation $\mathbb E \big[ {Z_j^T} \big]$}
During the transmission stage, the time interval for delivering the data package is directly related to both the file size and the transmission rate. In particular, the instantaneous transmission rate is given by
\begin{align} \label{Eq:Rate_Random_Variable}
R_{D} = B{\log _2}\left( {1 + {\gamma _D}} \right)
\end{align}
where $\gamma _{{D}} = {{{p_A}h {d}^{ - {\alpha}}}}/{{{\sigma ^2}}}$ is the signal-to-noise ratio (SNR) at destination node, with $h$ and ${\sigma ^2}$ representing the effect of Rayleigh fading and background noise, respectively. In addition, $d$ is the distance between the aggregator and destination node. We note that the transmission rate $R_{D}$ is a random variable under the effect of Rayleigh fading. As such, the expectation of transmission time for the package originally generated by source $S_j$ is as shown in Theorem \ref{Theorem:Exp_Transmission_Time_TQ}.

\begin{theorem} \label{Theorem:Exp_Transmission_Time_TQ}
The expected transmission time for delivering a data package from source $S_j$ is given as
\begin{align} \label{Eq:Transsmission_Time_Exp_Sj}
\mathbb E \big[ {Z_j^T} \big] =  {\frac{{{\xi _j}{\sigma ^2 d^\alpha }}}{p_A} \! \int_0^\infty \!\!\!\! \exp\! \left( {\frac{{{\xi _j}}}{t} + \frac{{ {{\rm{1}} - {\rm{exp}}\left( {\frac{{{\xi _j}}}{t}} \right)} }}{{{p_A} \sigma^{-2} {d^{ - \alpha }}}}} \right) \frac{dt}{t}}
\end{align}
where ${\xi _j} = {{{{\tilde C}_j}\ln 2}}/{B}$.
\end{theorem}
\begin{IEEEproof}
The proof is given in Appendix \ref{Pro:Exp_Transmission_Time_TQ}.
\end{IEEEproof}

\subsection{Computing the Expectation $\mathbb E \big[ {W_j^T} \big]$}
For the transmission subsystem, we can model it with a G/G/1 FCFS queueing system by considering that the inter-arrival time and service time follow different general distributions. The exact analytical results (closed-form or numerical) are usually unavailable for the G/G/1 queue, especially for the case where the distribution about the arrival or departure is unknown\cite{Queueing_Systems_1976}. For our formulated transmission queue following a priority M/G/1 processing queue, it is impracticable to obtain the distribution of the inter-arrival and directly analyze the expectation of the waiting time for packages. In this light, we resort to implementing the principle of maximum entropy and get an information theoretic approximation of $\mathbb E \big[ {W_j^T} \big]$.

\begin{theorem} \label{Theorem:Exp_Waiting_Time_TQ}
In the transmission queue, the expectation $\mathbb E \big[ {W_j^T} \big]$ can be mathematically approximated by the following
\begin{align}   \label{Eq:Waiiting_Time_Transmission_Q_Fin}
\mathbb E \big[ {W_j^T} \big] \thickapprox \frac{ {\mu _j} {{{\left( {\sum_{j = 1}^J {{\lambda _j}} \mathbb E \big[ {Z_j^T} \big] } \right)}^{\rm{2}}}}}{{\sum_{i = 1}^J {{\lambda _i}{\mu _i}} \left( {1 - \sum_{j = 1}^J {{\lambda _j}}\mathbb  E \big[  {Z_j^T} \big] } \right)}}
\end{align}
where ${\mu _1} =1$ and ${\mu _j}$ is given in (\ref{Eq:Waiiting_Time_Transmission_Q_Ratio}), $\forall j \in \{2,3,\cdots, J\}$, which represents the ratio of the expectation of waiting time for packages from source $S_j$ to that for packages from source $1$ in the transmission queue, i.e., ${\mu _j}={\mathbb E\big[ {W_j^T} \big]}/{\mathbb E\big[ {W_1^T}\big]}$.
\end{theorem}
\begin{IEEEproof}
The proof is given in Appendix \ref{Pro:Exp_Waiting_Time_TQ}.
\end{IEEEproof}

\section{Simulation Results and Discussions}

\begin{figure} [!t]
\centering
  \subfigure[] \leavevmode \epsfxsize=2.6in  \epsfbox{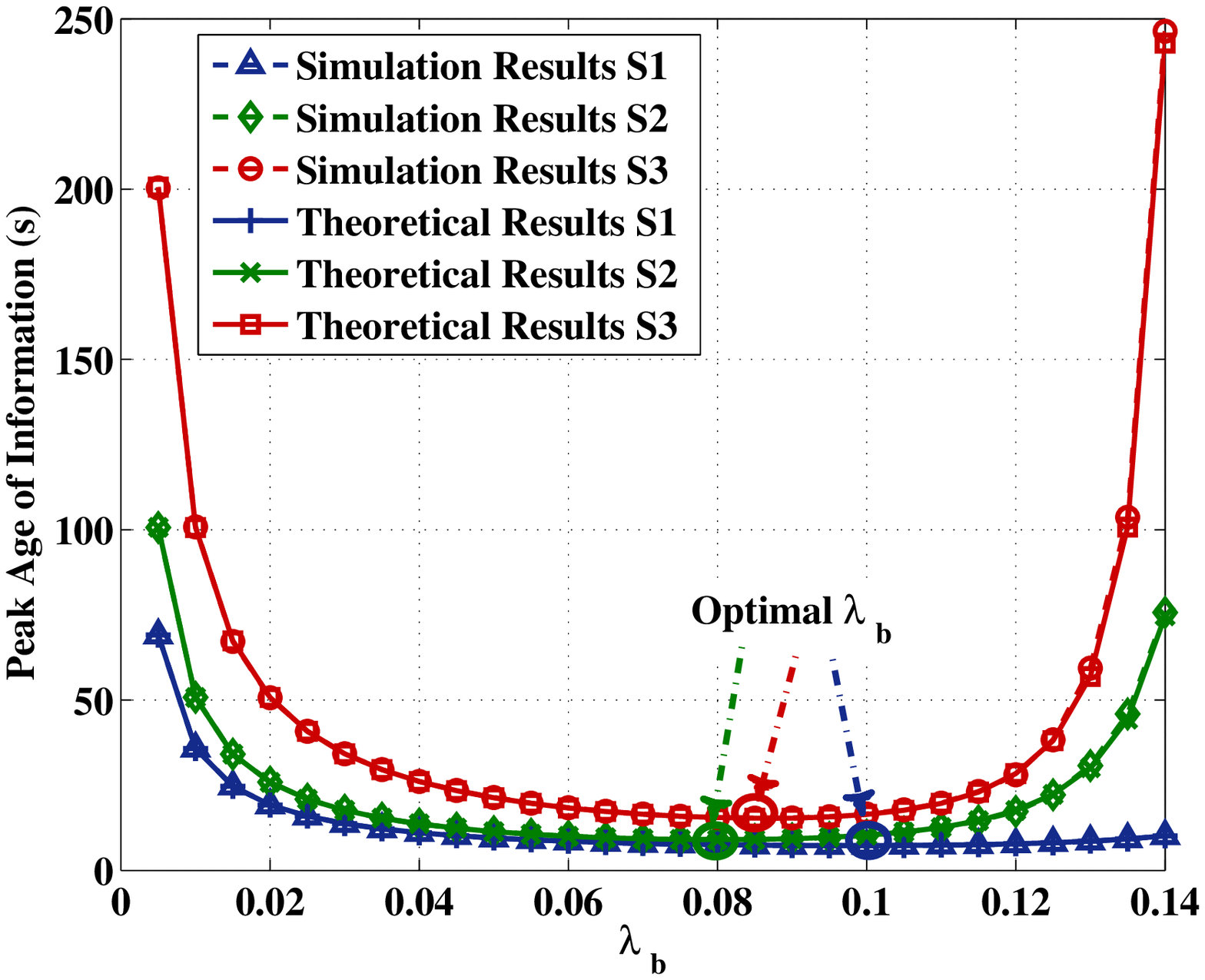}
  \subfigure[] \leavevmode \epsfxsize=2.6in  \epsfbox{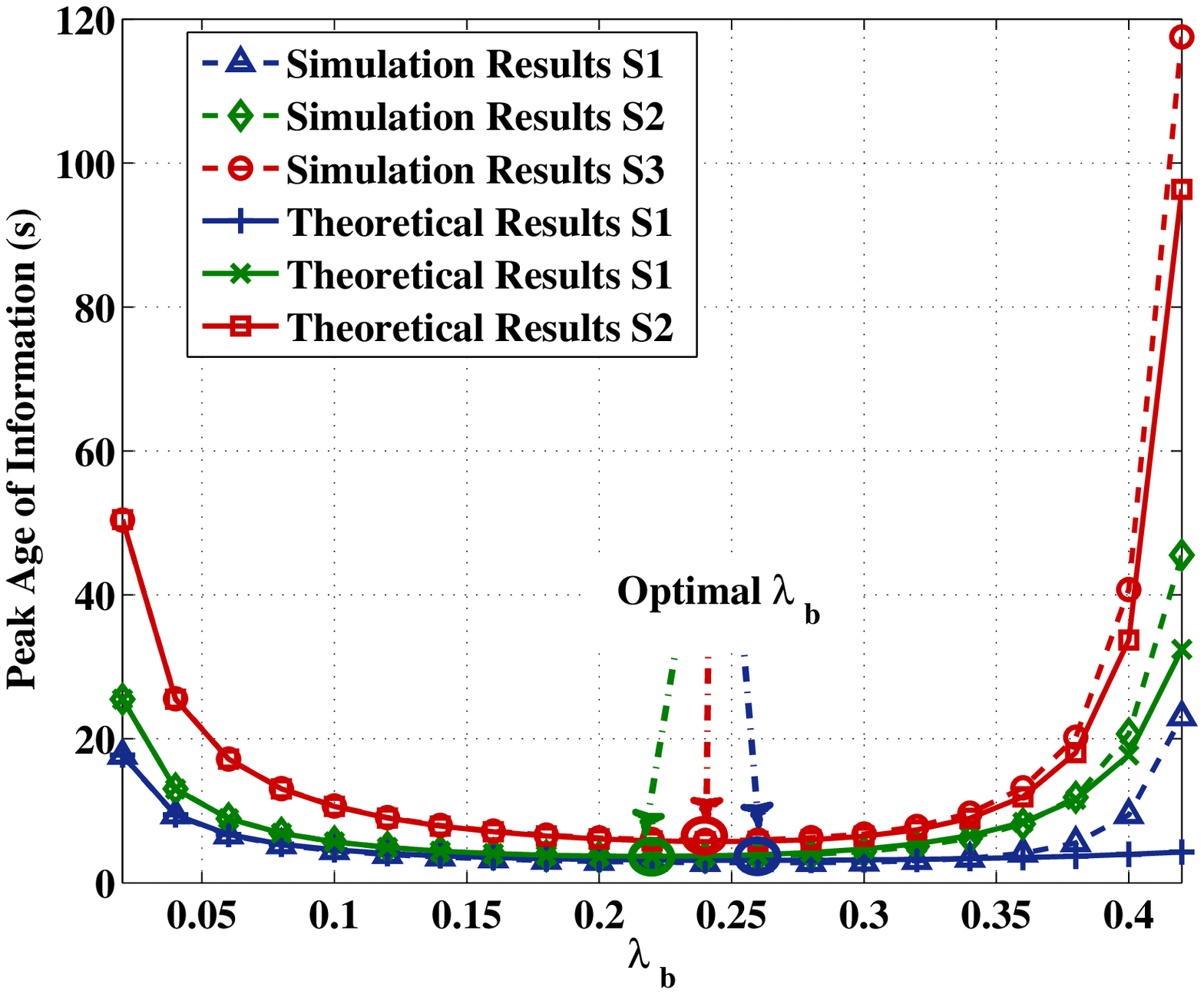}
\centering \caption{Average PAoI for packages from different sources, where the equivalent processing rate is: (a) 50 Mbits/s; (b) 150 Mbits/s.} \label{Fig:Theoretic_vs_Simulation}
\end{figure}

In this section, we verify the accuracy of our analysis via simulations. Specifically, we consider there are three sources and set the arrival rate of data packages from source $S_j$ to be $j\lambda _b$, i.e., $\lambda _j = j\lambda _b$, where $\lambda _b$ can be regarded as the ``basic" arrival rate belonging to $\left[ {0.005, 0.14} \right]$  with the units packages/second. The sizes of original and processed packages are set as $\{C_1, C_2, C_3\}=\{120, 35, 30\}$ Mbits and $\{\tilde C_1, \tilde C_2, \tilde C_3\}=\{20, 20, 20\}$ Mbits, respectively. For wireless transmission, the transmission power $p_A$ is 100 mW, distance $d$ is 200 m, transmission bandwidth $B$ is 1 MHz, and AWGN power density is -174 dBm/Hz. Meanwhile, each individual simulation result is obtained by averaging over PAoI for totally $10^6$ data packages.

Fig. \ref{Fig:Theoretic_vs_Simulation} (a) compares the simulation and theoretical results on the average PAoI for different sources, where the equivalent processing rate is $\tau_1 = \tau_2= \tau_2=50 $ Mbits/s, and the data processing time dominates transmission time, i.e., $\mathbb E \big[ {Z^P} \big] =  1.13$ s and $\mathbb E\big[ {{Z^T}} \big] = 0.371$ s. The results show a close match for all sources, which validate the our mathematical analysis. We note that even in the high traffic load region, e.g., $\lambda _b \ge 0.13$, the average PAoI for packages from source 1 is still kept low, i.e., about 8.1 s, while that for source 2 and 3 are about 22.3 s and 38.1 s, respectively. This is due to the fact that the priority based processing subsystem will try its best to first provide required service to packages with the highest priority even in the case where the resource is not enough to provide good service to all packages. In other words, the average PAoI for the packages with the lowest priority will first suffer significant performance degradation as the traffic load becomes heavy, while the state corresponding to packages with the highest priority is always kept in a stable region. Hence, to make the whole system working in a stable state it is necessary to properly control the package arrival rates for all sources.

To see what happens when the data transmission time is comparable to processing time, we set the equivalent processing rate to $150 $ Mbits/s for all packages, i.e., $\mathbb E \big[{Z^P} \big] = 0.378$ and $\mathbb E\big[ {{Z^T}} \big] = 0.371$. The simulation and theoretical results are illustrated in Fig. \ref{Fig:Theoretic_vs_Simulation} (b) where the similar observation can also be made as shown in Fig. \ref{Fig:Theoretic_vs_Simulation} (a). While we note that the difference between the simulation results and theoretical results is higher than that in Fig. \ref{Fig:Theoretic_vs_Simulation} (a), the shape of the average PAoI and the optimal basic arrival rate $\lambda _b^ *$ can still be well captured by our theoretical analysis. Particularly, in Fig. \ref{Fig:Theoretic_vs_Simulation} (b), for both the obtained simulation results and theoretical results, the optimal basic arrival rates for each individual source is identical, i.e., $\lambda _{b,1}^ * = 0.26$, $\lambda _{b,2}^ * = 0.22$, and $\lambda _{b,3}^ * = 0.24$.\footnote{According to the theoretical results, $\Delta _1$ first decreases with respect to the basic arrival rate $\lambda _b$ and then increases. The minimal of $\Delta _1$ is 3.12 for $\lambda _{b,1}^ * = 0.26$ while $\Delta _1 = 4.28$ when $\lambda _{b,1}$ is $0.42$.} Therefore, it's plausible to qualitatively obtain the optimal data arrival rates for distinct resources with our theoretical analysis presented in this section, when the processing time is dominating or comparable with the transmission time. However, the accurate analysis for more general cases is essentially complicated and hence left for our future work.

\section{Conclusion}
In this article, we took a fresh look at studying the information freshness in IoT networks with multiple data sources. Employing a realistic system model that allows the collected raw data to be preprocessed before being forwarded to the destination node, we modeled the system as a tandem queue and derived an analytical expression for the peak age of information. Simulations showed that our analyses are accurate when the processing time is dominating or comparable with the transmission time. There are many interesting extensions for this work, one of which is further investigating the optimal strategy on controlling the update frequency to make the delivered information as fresh as possible.

\appendices
\section{Proof of Theorem \ref{Theorem:Exp_Waiting_Time_PQ}} \label{Pro:Exp_Waiting_Time_PQ}
\begin{IEEEproof}
We denote the average number of packages with priority $j$ in processing queue by $\mathbb E [{N_{j,Q}^P}]$ and the expectation of the remaining processing time of a package in service by $\mathbb E [{Z_R^P}]$. The expectation $\mathbb E [ {W_1^P} ]$ can be expressed as
\begin{align} \label{Eq:Waiting_Time_S1}
\nonumber \mathbb E \big[ {W_1^P} \big] &={P_{{A_P},B}} \mathbb E \big[ {Z_R^P} \big] +\left(\! {1{ - }{P_{{A_P},B}}} \!\right) \!\cdot 0{\rm{ + }} \mathbb E \big[ {N_{1,Q}^P} \big] \mathbb E \big[ {Z_1^P} \big] \\
%\nonumber &{\rm{ = }}{P_{{A_P},B}} \mathbb E\left( {Z_R^P} \right){\rm{ + }}{\lambda _1} \mathbb E\left( {W_1^P} \right)E\left( {Z_1^P} \right) \\
&=\frac{{{P_{{A_P},B}}E \big[ {Z_R^P} \big] }}{{1{\rm{ - }}{\lambda _1}E \big[ {Z_1^P} \big] }}{\rm{ = }}\frac{{{P_{{A_P},B}}E \big[ {Z_R^P} \big] }}{{1{\rm{ - }}{\rho _1}}}
\end{align}
where ${P_{{A_P},B}}$ is expressed in Eq. (\ref{Eq:Busy_Probability_AP}) and ${\rho _1}$ denotes the load in the processing subsystem caused by packages with priority 1. Similarly, for $j>1$ we have
\begin{align} \label{Eq:Waiting_Time_SJ}
\nonumber \mathbb E \big[ {W_j^P} \big] {\rm{ = }}&{P_{{A_P},B}} \mathbb E \big[ {Z_R^P} \big] {\rm{ + }}\left( {1{\rm{ - }}{P_{{A_P},B}}} \right) \cdot 0{\rm{ + }} \\ \nonumber
&\sum\nolimits_{i = 1}^j {\mathbb E \big[ {N_{i,Q}^P} \big] \mathbb E \big[ {Z_i^P} \big] }+\sum\nolimits_{i = 1}^{j - 1} {{\lambda _i}\mathbb E \big[ {W_j^P} \big] \mathbb E \big[  {Z_i^P}  \big] }\\
%\nonumber {\rm{ = }}&{P_{{A_P},B}}\mathbb E\left( {Z_R^P} \right){\rm{ + }}\sum\limits_{i = 1}^j {{\lambda _i}\mathbb E\left( {W_i^P} \right)\mathbb E\left( {Z_i^P} \right)}  + \\ \nonumber
%& \mathbb E\left( {W_j^P} \right)\sum\limits_{i = 1}^{j - 1} {{\lambda _i}\mathbb E\left( {Z_i^P} \right)} \\
{\rm{ = }}&\frac{{{P_{{A_P},B}}\mathbb E \big[ {Z_R^P} \big]  + \sum_{i = 1}^{j - 1} {{\rho _i}\mathbb E \big[ {W_i^P} \big] } }}{{1 - \sum_{i = 1}^j {{\rho _i}} }}
\end{align}
Based on Eq. (\ref{Eq:Waiting_Time_S1}) and (\ref{Eq:Waiting_Time_SJ}) we have
\begin{align} \label{Eq:Waiting_Time_S2}
\nonumber \mathbb E \big[ {W_2^P} \big] {\rm{ = }}&\frac{{{P_{{A_P},B}}\mathbb E \big[ {Z_R^P} \big] + {\rho _i}\frac{{{P_{{A_P},B}}\mathbb E [ {Z_R^P} ] }}{{1{\rm{ - }}{\rho _1}}}}}{{1 - \sum_{i = 1}^2 {{\rho _i}} }} \\
{\rm{ = }}& \frac{{{P_{{A_P},B}}\mathbb E \big[ {Z_R^P} \big] }}{{\left( {1 - \sum_{i = 1}^2 {{\rho _i}} } \right)\left( {1{\rm{ - }}{\rho _1}} \right)}}
\end{align}
and, for $\forall j >2$,
\begin{align} \label{Eq:Waiting_Time_Recurrence}
\nonumber & \left( {1  \! -  \! \sum\nolimits_{i = 1}^j {{\rho _i}} } \right)\mathbb E \big[ {W_j^P} \big] \! = \! {P_{{A_P},B}}\mathbb E \big[ {Z_R^P} \big]  \! +  \! \sum\nolimits_{i = 1}^{j - 1} {{\rho _i}\mathbb E \big[ {W_i^P} \big] }  \\ \nonumber
&={P_{{A_P},B}}\mathbb E \big[ {Z_R^P} \big] + \sum\nolimits_{i = 1}^{j - 2} {{\rho _i}\mathbb E \big[ {W_i^P} \big] }  + {\rho_{j - 1}}\mathbb E \big[ {W_{j - 1}^P} \big]  \\
%\nonumber &= \left( {1 - \sum\limits_{i = 1}^{j - 1} {{\rho _i}} } \right)\mathbb E\left( {W_{j - 1}^P} \right) + {\rho _{j - 1}}\mathbb E\left( {W_{j - 1}^P} \right)   \\
&= \left( {1 - \sum\nolimits_{i = 1}^{j - 2} {{\rho _i}} } \right)\mathbb E \big[ {W_{j - 1}^P} \big].
\end{align}
Substituting (\ref{Eq:Waiting_Time_S2}) into the recursion formula in (\ref{Eq:Waiting_Time_Recurrence}), we can express the expectation $\mathbb E\big[ {W_j^P} \big]$ as
\begin{align} \label{Eq:Processing_Waiting_Time_Fin}
\mathbb E \big[ {W_j^P} \big] =
\begin{cases}
\frac{{{P_{{A_P},B}}\mathbb E [ {Z_R^P} ] }}{{1{\rm{ - }}{\rho _1}}}, &j=1 \\
\frac{{{P_{{A_P},B}}\mathbb E [ {Z_R^P} ] }}{{\left( {1 - \sum\nolimits_{j = 1}^{J} {{\rho _i}} } \right)\left( {1 - \sum\nolimits_{j = 1}^{J-1} {{\rho _i}} } \right)}}, & j >1
\end{cases}
\end{align}
where $\mathbb E\big[{Z_R^P} \big]$ is the expectation of the remaining processing time of a package in service.
By applying the renewal-reward theory \cite{Queueing_Performance}, we have
\begin{align} \label{Eq:Remaining Processing_Time_Org}
\mathbb E \big[ {Z_R^P} \big] = {{\mathbb E \Big[ \left( {{Z^P}} \right)^2 \Big] }}/{{2\mathbb E \big[ {{Z^P}} \big] }}
\end{align}
where
\begin{align} \label{Eq:Processing_Time_1M}
\mathbb E \big[ {{Z^P}} \big] = \sum\nolimits_{j = 1}^J {  {\frac{{{\lambda _j}}}{{\sum_{i = 1}^J {{\lambda _i}} }}\mathbb E \big[ {Z_j^P} \big] } }  = \frac{{\sum_{j = 1}^J {{\rho _j}} }}{{\sum_{j = 1}^J {{\lambda _i}} }}
\end{align}
and
\begin{align} \label{Eq:Processing_Time_2M}
\mathbb E\Big[ {{{\left(\! {{Z^P}}\! \right)}^2}} \Big] \! = \! \sum\nolimits_{j = 1}^J \!{  {\frac{{{\lambda _j}}}{{\sum_{i = 1}^J {{\lambda _i}} }}{{\left( {\mathbb E \big[ {Z_j^P} \big] } \right)}^2}}  } \! = \! \frac{{\sum_{j = 1}^J \! {{{{{ {{\rho^2_j}} }}}}/{{{\lambda _j}}}} }}{{\sum_{j = 1}^J \! {{\lambda _i}} }}.
\end{align}

Finally, combining from Eq. (\ref{Eq:Processing_Waiting_Time_Fin}) to (\ref{Eq:Processing_Time_2M}), we can draw the conclusion shown in Theorem
\ref{Theorem:Exp_Waiting_Time_PQ}.
\end{IEEEproof}

\section{Proof of Theorem \ref{Theorem:Exp_Transmission_Time_TQ}} \label{Pro:Exp_Transmission_Time_TQ}
\begin{IEEEproof}
According to (\ref{Eq:Rate_Random_Variable}), the transmission time of one package from the aggregator to destination node is
\begin{align} \label{Eq:Time_Random_Variable}
Z_{\tilde{C}}^T = \frac{{{\tilde{C}}}}{{R_D}} = \frac{{{\tilde{C}}\ln 2}}{{B}}\frac{1}{{\ln \left( {1 + \frac{{{p_A}h{d}^{ - {\alpha}}}}{{{\sigma ^2}}}} \right)}}
\end{align}
where $\tilde{C} \in \{\tilde{C}_1, \tilde{C}_2, \cdots, \tilde{C}_J\}$ denotes the size of the concerned package. Note that $Z_{\tilde{C}}^T$ is a random variable due to the random channel gain. Moreover, $Z_{\tilde{C}}^T$ monotonically decreases with respect to the channel gain $h$
with the expression as
\begin{align}
h = \frac{{{\sigma ^2}( {{\rm{exp}}( {\frac{{{\tilde{C}}\ln 2}}{{B}}\frac{1}{{Z_{\tilde{C}}^T}}} ) - 1} )}}{{{p_A}{d}^{ - {\alpha}}}} = f( {Z_{\tilde{C}}^T} )
\end{align}
where $f( {Z_{\tilde{C}}^T} )$ is the function inversely mapping from $Z_{\tilde{C}}^T$ to $h$. Then, we obtain the probability density function of $Z_{\tilde{C}}^T$ as
\begin{align} \label{Eq:Time_PDF}
{f_{Z_{\tilde{C}}^T}}\left( t \right) &= - \exp \left( { - f\left( t \right)} \right)\frac {df\left( t \right)}{dt} \\ \nonumber
&= \frac{{\tilde C\ln 2{\sigma ^2}}}{{B{p_A}{{ {d}}^{ - {\alpha}}}}}\frac{{\exp \left( {\frac{{\tilde C\ln 2}}{{Bt}} + \frac{{{\sigma ^2}\left( {{\rm{1}} - {\rm{exp}}\left( {\frac{{\tilde C\ln 2}}{{Bt}}} \right)} \right)}}{{{p_A}{{ {d} }^{ - {\alpha}}}}}} \right)}}{{{t^2}}}.
%=  - \exp \left( { - \frac{{{\sigma ^2}\left( {{\rm{exp}}\left( {\frac{{\tilde C\ln 2}}{B}\frac{1}{t}} \right) - 1} \right)}}{{{p_S}{{\left( {d_A^D} \right)}^{ - {\alpha _A}}}}}} \right)\frac{{{\sigma ^2}{\rm{exp}}\left( {\frac{{\tilde C\ln 2}}{B}\frac{1}{t}} \right)}}{{{p_S}{{\left( {d_A^D} \right)}^{ - {\alpha _A}}}}}\frac{{\frac{{\tilde C\ln 2}}{B}}}{{ - {t^2}}} \\ \nonumber
\end{align}

As such, for packages originally generated from source $S_j$, the expectation of transmission time can be attained as
\begin{align} \label{Eq:Transsmission_Time_Exp_Sj_Def}
\mathbb E\big[ {Z_j^T} \big] = \mathbb E\Big[ {Z_{\tilde C}^T\left| {\tilde C{\rm{ = }}{{\tilde C}_j}} \right.} \Big] = \int_0^\infty  {t{f_{Z_{\tilde C}^T\left| {\tilde C{\rm{ = }}{{\tilde C}_j}} \right.}}\left( t \right)} dt.
\end{align}
Finally, by substituting (\ref{Eq:Time_PDF}) into (\ref{Eq:Transsmission_Time_Exp_Sj_Def}) we can draw the conclusion in Theorem \ref{Theorem:Exp_Transmission_Time_TQ}.
\end{IEEEproof}

\section{Proof of Theorem \ref{Theorem:Exp_Waiting_Time_TQ}} \label{Pro:Exp_Waiting_Time_TQ}
\begin{IEEEproof}
We adopt the principle of maximum entropy (PME) to derive an approximation of the expectation $\mathbb E\big[ {W_j^T} \big]$. The interested readers are suggested to refer \cite{EM_PL,EM_Q_3} for more details about PME and its applications for performance analysis in various types of queueing systems.

In the transmission queue, the expectation of the waiting time for a typical data package in the queue can be expressed
\begin{align}   \label{Eq:Waiiting_Time_Transmission_Q_No_Diff_Org}
\mathbb E\big[ {{W^T}} \big] \!  = \! \sum\nolimits_{j = 1}^J \! {P_j^T \mathbb E\big[ {W_j^T} \big]} \! \mathop  = \limits^{\left( a \right)} \! \sum\nolimits_{j = 1}^J \! \!{\frac{{{\lambda _j}}}{{\sum_{j = 1}^J \! {{\lambda _j}}}}\mathbb E\big[ {W_j^T} \big]}
\end{align}
where $P_j^T$ denotes the probability that there is one package arriving at the transmission queue originally from source $S_j$, $\mathbb E\big[ {W_j^T} \big]$ is the expectation of its waiting time, and (a) holds under the condition that the previous processing subsystem is stable, i.e., the arrivals are all processed on average. Moreover, from Theorem \ref{Theorem:Exp_Transmission_Time_TQ} we have that for a typical package, the average time spent in the transmission subsystem can be expressed as
\begin{align} \label{Eq:Transsmission_Time_Exp}
&\mathbb E\big[ {{Z^T}}\big]= \sum\nolimits_{j = 1}^J {\mathbb E\Big[ {Z_{\tilde C}^T\left| {\tilde C{\rm{ = }}{{\tilde C}_j}} \right.} \Big]P\left( {\tilde C{\rm{ = }}{{\tilde C}_j}} \right)} \\ \nonumber
&\mathop  = \limits^{\left( a \right)} \sum\nolimits_{j = 1}^J {\frac{{{\lambda _j}}}{{\sum_{i = 1}^J {{\lambda _i}} }}\mathbb E\Big[ {Z_{\tilde C}^T\left| {\tilde C{\rm{ = }}{{\tilde C}_j}} \right.} \Big]}  = \frac{{\sum_{j = 1}^J {{\lambda _j}\mathbb E\big[ {Z_j^T} \big]} }}{{\sum_{j = 1}^J {{\lambda _j}} }}
\end{align}
where ($a$) holds under the condition that the previous processing subsystem is stable, and $\mathbb E\big[ {Z_j^T} \big]$ is given in (\ref{Eq:Transsmission_Time_Exp_Sj}). Applying Little's law and combing the result with (\ref{Eq:Waiiting_Time_Transmission_Q_No_Diff_Org}) we furtjer have
\begin{align}   \label{Eq:Waiiting_Time_Transmission_Q_Org}
&\mathbb E\big[ {W^T} \big]= \frac {\mathbb E \big[N^T\big]}{\sum_{j = 1}^J {{\lambda _j}}} - \mathbb E\big[ {{Z^T}} \big] \\ \nonumber
&= \sum\nolimits_{j = 1}^J {\frac{{{\lambda _j}}}{{\sum_{j = 1}^J {{\lambda _j}} }}\mathbb E\big[ {W_j^T} \big]} = \sum\nolimits_{j = 1}^J {\frac{{{\lambda _j}{\mu _j}}}{{\sum_{j = 1}^J {{\lambda _j}} }}\mathbb E\big[W_1^T\big]}
\end{align}
where $\mathbb E \big[N^T\big]$ denotes the expectation of the total number of packages in the transmission subsystem, $\mathbb E\big[ {{Z^T}} \big]$ is given by (\ref{Eq:Transsmission_Time_Exp}), and $\mu _j$ represents the ratio ${\mathbb E\big[ {W_j^T} \big]}/{\mathbb E\big[ {W_1^T}\big]}, \forall j \in \{1, 2, \cdots, J\}$. According to Eq. (\ref{Eq:Waiiting_Time_Transmission_Q_Org}), we can obtain $\mathbb E\big[ {W_j^T} \big]$ if $\mathbb E \big[N^T\big]$ and $\mu _j, \forall j \in \{2, \cdots, J\}$, are derived.

As $N^T$ is an integer-value random variable, we use the PME to express its probability mass function as follows \cite{EM_PL}
\begin{align}   \label{Eq:PMF_Our}
P\left( {{N^T = n}} \right) &= \frac{1}{G}\exp \left( { - \sum\nolimits_{m = 1}^M {{\beta _m}{{\left( n \right)}^m}}} \right) \\ \nonumber
& \mathop  \approx \limits^{(a)} \frac{1}{G}\exp \left( { - {\beta _1 }n} \right), \forall n \in \{0, 1, 2, \cdots\}
\end{align}
where
\begin{align}   \label{Eq:PMF_Parameter_Z_Our}
G &= \sum\nolimits_{n = 0}^\infty  {\left( { - \sum\nolimits_{m = 1}^M {{\beta _m}{{\left( n \right)}^m}} } \right)}  \\ \nonumber
& \mathop  \approx \limits^{(b)} \sum\nolimits_{n = 0}^\infty  {\exp \left( { - {\beta _1}n} \right)}  = {\left( {1 - \exp \left( { - {\beta _1}} \right)} \right)^{ - 1}}.
\end{align}
Wherein, $\beta_m$ is the introduced Lagrangian multiplier associated with the $m$-th moment of the random variable $N^T$, while (a) and (b) hold due to the first moment approximation\footnote{Note that the accuracy of the approximation improves when more moments of $N^T$ are incorporated, giving rise to higher complexity. }.

By applying Little's law to the transmission queue and combining the result with (\ref{Eq:PMF_Our}) and (\ref{Eq:PMF_Parameter_Z_Our}) we have the following
\begin{align}   \label{Eq:PMF_Zero}
P\left( {{N^T = 0}} \right)  = 1-\rho ^T  \approx  \left( {1 - \exp \left( { - {\beta _1}} \right)} \right)
\end{align}
where $\rho ^T=\sum_{j = 1}^J {{\lambda _j}} \mathbb E\big[ {{Z_j^T}} \big]$ denotes the probability that the server is busy. Using the PME for another time, we have \cite{EM_PL}
\begin{align}   \label{Eq:Transmission_Q_User_Num}
{\mathbb E \big[N^T\big]} & \! \thickapprox \! \frac{{\partial \ln \left( {1 \!-\! \exp \left( { \! - {\beta _1}} \! \right)} \right)}}{{\partial {\beta _1}}} \!= \! \frac{{\sum_{j = 1}^J {{\lambda _j}} \mathbb E\big[ {{Z_j^T}} \big]}}{{1 - \sum_{j = 1}^J {{\lambda _j}} \mathbb E\big[ {{Z_j^T}}\big]}}.
\end{align}

Next, we analyze the ratio $\mu _j$. In the transmission queue, the waiting time of one arriving data package is related to the number and kinds of packages waiting in front of it, which are determined by the output of the previous processing queue and are extremely difficult to obtain. Recalling the analysis for $\mathbb E\big[W_j^P\big]$ in Appendix \ref{Pro:Exp_Waiting_Time_PQ}, we can derive an approximation of ratio $\mu _j, \forall j \in \{2,3,\cdots, J\},$ by considering that the packages in front of an arriving package in the transmission queue are proximately proportional to those in the processing queue, i.e., as shown in Eq. (\ref{Eq:Waiiting_Time_Transmission_Q_Ratio}). Finally, substituting (\ref{Eq:Transmission_Q_User_Num}) and (\ref{Eq:Waiiting_Time_Transmission_Q_Ratio}) into (\ref{Eq:Waiiting_Time_Transmission_Q_Org}) we can draw the conclusion.
\end{IEEEproof}

\ifCLASSOPTIONcompsoc
  \section*{Acknowledgments}
\else
  \section*{Acknowledgment}
\fi

% The authors would like to thank the Editor and the Reviewers for their precious time and efforts in reviewing this paper and providing helpful comments and suggestions.
This work is supported by the National Natural Science Foundation of China (61701372) and Talents Special Foundation of Northwest A\&F University (Z111021801).

\ifCLASSOPTIONcaptionsoff
  \newpage
\fi

%%%%%%%%%%%%%%%%%%%%%%%%%%%%%%%%%%%%%%%%%%%%%%%%%%%%%%%%%%%%%%%%%%%%%%%%%%%%%%%%%
%%%%%%%%%%%%%%%%%%%%%%%%%%%%%%%%%%%%%%%%%%%%%%%%%%%%%%%%%%%%%%%%%%%%%%%%%%%%%%%%%

\bibliographystyle{IEEEtran}
\bibliography{IEEEabrv,CUTD_Ref}

\end{document}